\documentclass{article}

\usepackage{microtype}
\usepackage{graphicx}
\usepackage{subfigure}
\usepackage{booktabs} 

\usepackage{hyperref}

\usepackage[accepted]{icml2019}

\usepackage{amssymb,amsmath,mathtools}
\usepackage{amsthm}

\newcommand{\dpar}[2]{\frac{\partial #1}{\partial #2}}

\newcommand{\tp}{^{\mathrm{T}}}  
\AtBeginDocument{\mathcode`\;=\numexpr\mathcode`\;-\string"4000\relax} 

\newcommand{\bphi}{\boldsymbol{\varphi}}

\newcommand{\muskern}{\kern-.15ex } 
\makeatletter
\newcommand\dynmark[1]{{\normalfont\bfseries\itshape
  \@tfor\next:=#1\do{\put@muskern\next}\/}}
\newcommand{\put@muskern}{\let\put@muskern\muskern}
\makeatother

\icmltitlerunning{User Curated Shaping of Expressive Performances}

\begin{document}

\twocolumn[
\icmltitle{User Curated Shaping of Expressive Performances}



\icmlsetsymbol{equal}{*}

\begin{icmlauthorlist}
\icmlauthor{Zhengshan Shi}{st}
\icmlauthor{Carlos Cancino-Chac\'{o}n}{of}
\icmlauthor{Gerhard Widmer}{of,jku}
\end{icmlauthorlist}

\icmlaffiliation{st}{CCRMA, Stanford University, USA}
\icmlaffiliation{of}{Austrian Research Institute for Artificial Intelligence, Vienna, Austria}
\icmlaffiliation{jku}{Johannes Kepler University Linz, Austria}

\icmlcorrespondingauthor{Zhengshan Shi}{kittyshi@ccrma.stanford.edu}
\icmlcorrespondingauthor{Carlos Cancino-Chac\'{o}n}{carlos.cancino@ofai.at}

\icmlkeywords{Machine Learning, ICML}

\vskip 0.3in
]



\printAffiliationsAndNotice{}  

\begin{abstract}
Musicians produce individualized, expressive performances by manipulating parameters such as dynamics, tempo and articulation.
This manipulation of expressive parameters is informed by elements of score information such as pitch, meter, and tempo and  dynamics markings (among others).
In this paper we present an interactive interface that gives users the opportunity to explore the relationship between structural elements of a score and expressive parameters.
This interface draws on the basis function models, a data-driven framework for expressive performance.
In this framework, expressive parameters are modeled as a function of score features, i.e., numerical encodings of specific aspects of a musical score, using neural networks.
With the proposed interface, users are able to weight the contribution of individual score features and understand how an expressive performance is constructed.


\end{abstract}

\section{Introduction}
\label{sec:introduction}

The way a piece of music is performed expressively constitutes a very important aspect of our enjoyment of the music.
In Western art music, performers convey expression in their performances through variations in expressive dimensions such as tempo, dynamics and articulation, among others.

While most computational models of expressive performance allow for modeling only a single performance strategy, musicians can interpret a piece of music with a wide variety of stylistic and expressive inflections~\cite{Kirke:2013th,CancinoChacon:2018po}.
Some computational models allow users to control global characteristics of the performance (like tempo and dynamics) in real time~\cite{dixon2005air,Chew:2006:ERC:1178723.1178744,baba2010virtualphilharmony}.
In this work we present a prototype of a system that allows users to generate individualized piano performances by weighting the contribution of individual aspects of the musical score to the overall performance.


The rest of this paper is structured as follows:
Section \ref{sec:basis_function_models} provides a brief overview of the basis function models.
Section \ref{sec:user_controlled} describes the proposed extension to the basis function models to allow the user to weight the contribution of individual aspects of the score to shape expressive performances.
Finally, the paper is concluded in Section \ref{sec:conclusions}.

\section{Basis Function Models}
\label{sec:basis_function_models}

The basis function models are a data-driven framework for modeling musical expressive performance of notated music~\cite{Grachten:2012hk,CancinoChacon:2016wi}.
In this framework, numerical representations of expressive dimensions such as tempo and dynamics (which we refer to as \emph{expressive parameters}) are modeled as function of \emph{score basis functions}: numerical encodings of structural aspects of a musical score.
These aspects include low-level notated features such as pitch and metrical information, as well as music theoretic features and cognitively motivated features.
More formally, an expressive parameter can be written as $y_i = f(\bphi_i)$,
where $\bphi_i$ is a vector of basis functions evaluated on score element $x_i$ (e.g., a note or a position in the score, which we refer to as \emph{score onset}) and $f(\cdot)$ is (non-linear) function (the output of a neural network, as described below).
For a thorough description of the basis function models, see \cite{Cancino2018:th}.

\subsection{Representing Performance Information}

In order to capture the sequential nature of music, we divide the performance information into onset-wise and note-wise parameters.
Onset-wise parameters capture aspects of the performance with respect to the corresponding temporal (score) position, while note-wise features capture aspects of the performance of each note:

\subsubsection{Onset-wise parameters}
\begin{enumerate}
\item \textbf{MIDI velocity trend} (vt). Maximal MIDI velocity at each score onset
\item \textbf{Log Beat Period Ratio} (lbpr). Logarithm of the beat period, i.e., the time interval between consecutive beat grids, divided by the average beat period of the piece. 
\end{enumerate}
\subsubsection{Note-wise parameters}
\begin{enumerate}
\item \textbf{MIDI velocity deviations} (vd). The deviation of the MIDI velocity for each note from the trend.
\item \textbf{Timing} (tim). Onset deviations of the individual notes from the grid established by the local beat period.
\item \textbf{Articulation} (art). Logarithm of the ratio of the actual duration of a performed note to its reference (notated) duration according to the local beat period.
\end{enumerate}

These parameters are then standardized per piece to be zero-mean and unit variance.

%
%


\subsection{Modeling Expressive Performance}
We use bi-directional LSTMs to model onset-wise as well as note-wise parameters, given their sequential nature.
The input of the networks for predicting onset-wise parameters are the basis functions evaluated for each score onset, while the input of the networks for note-wise parameters are the basis functions evaluated for every note.


The models are trained in a supervised fashion to minimize the reconstruction error on the Magaloff/Chopin \cite{Flossmann:2011ty} and Zeilinger/Beethoven \cite{CancinoChacon:2017ht} datasets.
These datasets consists of recordings of piano music performed on computer controlled B\"osendorfer grand pianos, which have been aligned to their scores.

\section{User-controlled Basis Function Models}\label{sec:user_controlled}

In order to allow the users to explore and adjust the contribution of individual score features, we need first to compute the contribution of a feature to the output of the model.
A way to do so is to define a locally-linear approximation of the output of the neural networks modeling each expressive parameter as follows
\begin{equation}
\tilde{y}_i = c + \left({\dpar{}{\bphi}f(\bphi_{*})}\right)\tp (\bphi_i - \bphi_{*}), 
\end{equation}
where $\tilde{y}$ is the approximated value of the expressive parameter for score element $x_i$, $c$ is a user defined constant value (e.g.~the average lbpr or MIDI velocity of the piece), $\bphi_i$ is the vector of basis functions evaluated for score element $x_i$ and ${\dpar{}{\bphi}f(\bphi_{*})}$ is the gradient of $f$ with respect to $\bphi$ evaluated in $\bphi_*$.
We can naturally extend this locally-linear approximation to onset-wise models, by constructing a \emph{temporal} Jacobian matrix, in which its $ij$-th element can be interpreted as the ``contribution'' of the $j$-th basis function (e.g., the pitch, the inter-onset-interval, etc) to the performance of the $i$-th score onset.

\subsection{Interactive Interface}
\begin{figure}
  \centering
  \includegraphics[width=\linewidth]{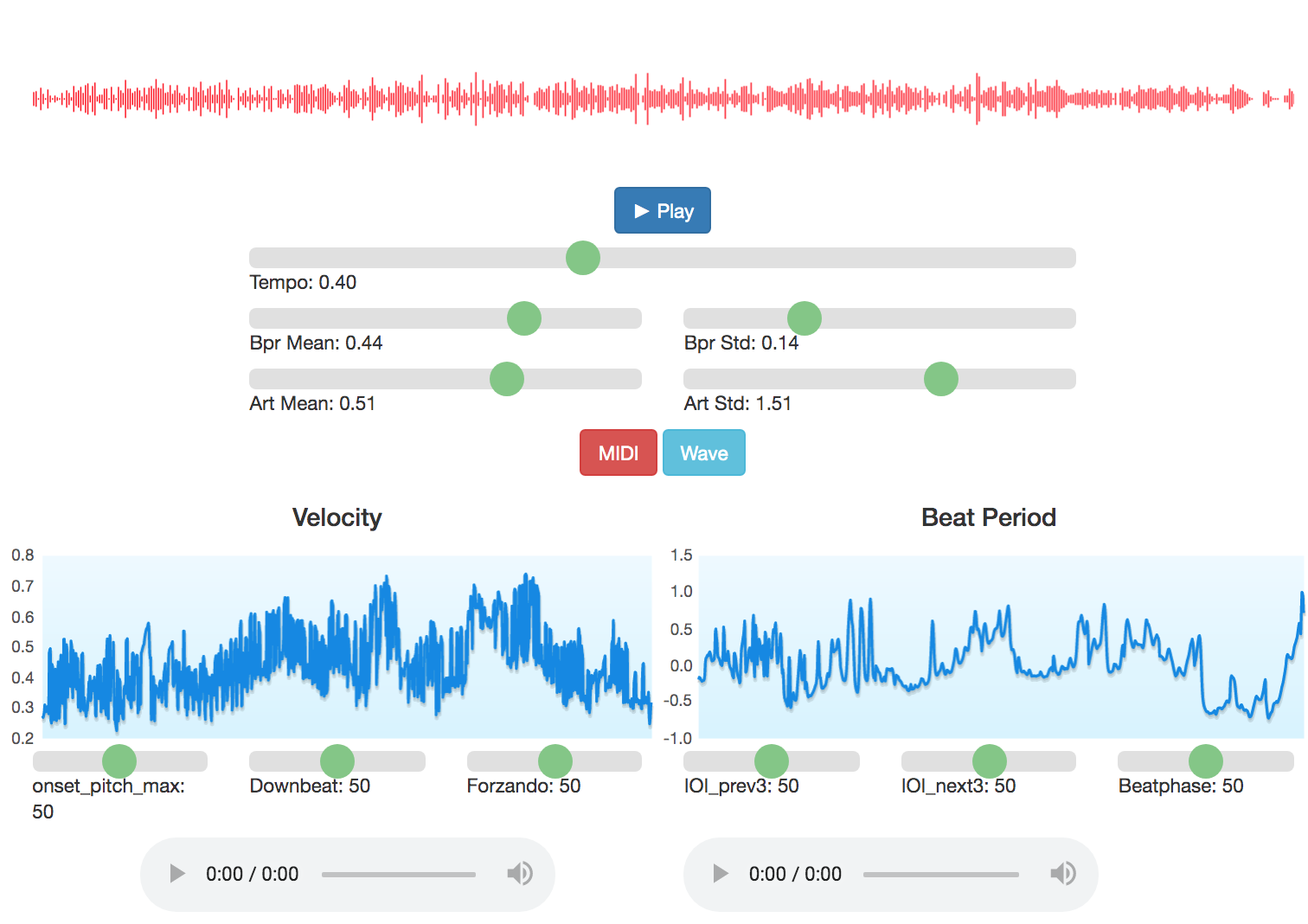}
  \caption{The user interface where the waveform of the predicted performance is displayed. Sliders are provided for the user to shape the performance. Curves indicating expressive parameters are updated as the user change the sliders. }
  \label{fig:scanner}
\end{figure}

The interface allows users to explore the contribution of individual score descriptors (e.g., the velocity on downbeats, the timing surrounding a beat phase) by adjusting the scaling of each column of the temporal Jacobian matrix. Curves indicating velocity and beat period will be updated to visualize the changes. The onset-wise and note-wise parameters will be calculated with the locally-linear approximation, and a new performance will be rendered and displayed for listening. The users will also be able to indicate their preference on overall tempo and articulation of the piece by adjusting the mean and standard deviation. In this way, they can shape the way a performance is rendered, while exploring the contribution of different musical dimensions (Figure \ref{fig:scanner}).


\section{Conclusions}\label{sec:conclusions}
In this paper we have presented a prototype of an interface that allows users to explore the contribution of individual score descriptors to the expressiveness of the performance.
Such an interface could have potential pedagogical applications: users can interactively explore the complex patterns through which score features contribute to the overall expressiveness, while at the same time allowing for creating personalized interpretation, as the performer gives more importance to certain parameters. 


\section*{Acknowledgements}
This research has received funding from the European Research Council (ERC) under the European Union's Horizon 2020 research and innovation programme under grant agreement No. 670035 (project ``Con Espressione").

\bibliography{bib_cc}
\bibliographystyle{icml2019}


\end{document}